\documentclass[singlespacing]{elsart}

\usepackage{epsfig}

\usepackage{amssymb}

\begin{document}

\begin{frontmatter}

\title{Unusual Behavior of Antiferromagnetic Superconductors in Low Magnetic Fields}

\author{Krzysztof Rogacki}
\ead{rogacki@int.pan.wroc.pl}

\address{Institut of Low temperature and Structure Research, Polish Academy of Sciences,
50-950 Wroclaw, Poland}

\begin{abstract}
In this article, we examine the superconducting properties of low-
and high-$T_c$ magnetic superconductors in magnetic fields close
to the first penetration field.  Attention is paid to the
properties that relate to the interactions between
antiferromagnetism and superconductivity.  It is suggested that
several features characterizing
the interplay between magnetic and superconducting subsystems in
low-$T_c$ superconductors can also be present in high-$T_c$
materials, however, they have not been observed for any
non-substituted antiferromagnetic superconductors of the Y123 type.
For the
Gd$_{1+x}$Ba$_{2-x}$Cu$_3$O$_{7-\delta}$ compound, a peak in the
temperature dependence of the ac
susceptibility has been found for $x = 0.2$ near the
N\'{e}el temperature of the Gd sublattice.  This peak is
attributed to the suppression of superconducting persistent
currents due to the pair breaking effect that results from the
enhanced magnetic fluctuations in the vicinity of the phase
transition temperature.  This
observation indicates that the interaction between magnetic and
conducting electrons is present for the composition with $x =
0.2$, where magnetism is enhanced and superconductivity
diminished.
\end{abstract}

\begin{keyword}
Magnetic superconductors, First penetration field, Vortex dynamics,
High-$T_c$ superconductors
\PACS 74.25.Ha \sep 74.60.Ec \sep 74.60.Ge \sep 74.70.Dd \sep 74.72.Jt
\end{keyword}

\end{frontmatter}

\section{Introduction}
\label{Introd}Magnetism and superconductivity are two forms of
long-range order that may exist in a material at appropriate low
temperatures.  When both are present and the singlet state paring
occures, they compete with one another,
and many authors have wondered under what conditions they
might coexist.  The answer has been emerging in recent years, but
the story began many years ago with theoretical works by Ginzburg
\cite{Ginzburg'1956}, Abrikosov and Gorkov
\cite{Abrikosov'1960}, and experimental studies by Matthias and
co-workers that intensively examined the solid solutions
La$_{1-x}$RE$_x$ \cite{Matthias'1958} and RERu$_2$
\cite{{Matthias'1958},{Matthias'1958a}} (RE =
magnetic rare earth ions). In La$_{1-x}$RE$_x$, superconductivity
was destroyed by relatively small amounts of paramagnetic ions
($\simeq 1$ at.\%), well before the long-range magnetic order
appeared. These experiments showed that superconductivity was
suppressed by the so-called $s$-$f$ exchange interaction between
conduction electrons and localized f-shell spins
\cite{{Matthias'1958},{Suhl'1959}}.

After many years of extensive studies of alloys and disordered
compounds with paramagnetic impurities it became clear that
long-range magnetic order and singlet state superconductivity may
coexist only in weak limit of the $s$-$f$ exchange interaction
\cite{Maple'1976}. One of the simplest ways to fulfil this
requirement consists in the spatial separation of conduction and
magnetic electrons.  This idea was first realized along with the
discovery of ternary compounds like RERh$_4$B$_4$ and
the REMo$_6$S$_8$ Chevrel phases \cite{Super.Tern.Comp.'1982}.
Most of these compounds become superconducting below a
temperature $T_c$ despite the presence of a relatively
large amount of the rare earth magnetic ions ($\simeq 11$ at.\%
for RERh$_4$B$_4$). Moreover, many of them show long-range
magnetic order in the superconducting state. For
ferromagnetic (FM) order, superconductivity can be preserved
only in a very narrow temperature range, just below the Curie
temperature.  However, when the magnetic order is
antiferromagnetic (AFM), superconductivity remains below the
Neel temperature, $T_N$, down to
the lowest investigated temperatures.  The main reason for the
observed coexistence of superconductivity and long-range
antiferromagnetism seems to be the particularly weak exchange
interaction due to the partial separation of the conduction $4d$
electrons in $k$-space from the localized $4f$ electrons of the
rare earth ions due to the cluster structure of these compounds.

In layered low-$T_c$ RENi$_2$B$_2$C superconductors
\cite{{Gupta'1997},{Schmidt'1998},{Canfield'1998},{Muller'2001}},
the spatial separation between conduction electrons, mostly Ni 3d
\cite{Ref.{Muller'2001}}, and magnetic electrons, RE 4f, is no
longer clear.  However, the weakness of the exchange interaction
and, therefore, the presence of coexistence may still be
understood. In these compounds, where RE = Dy, Ho, Er, and Tm,
the hybridization of 4$f$ and conduction electrons is weak and a
partial separation between those electrons appears in momentum
space. For RE = Pr, Nd, Pm, Sm, Eu, Gd, and Tb, the absence of
superconductivity involves several reasons which are beyond the
scope of this article and are reviewed elsewhere
\cite{Ref.{Muller'2001}a}.

In the case of layered high-$T_c$ REBa$_2$Cu$_3$O$_7$ (RE123)
superconductors, the coexistence between long-range
antiferromagnetism and superconductivity seems to be more puzzling.
In these compounds the magnetic RE ions are located between the
double CuO$_2$ planes that are responsible for superconductivity.
Here, as for the layered low-$T_c$ superconductors, the long-range
AFM order of the RE ions is realized via the RKKY and/or
superexchange interactions \cite{Maple'1989}.  For the RE123
materials, however, a modified exchange mechanism is necessary
because the RE magnetic moments and the conduction electrons interact
very weakly \cite{{Liu'1988},{Chattopadhyay'1988}}. An additional
reason that the coexistence of AFM order and superconductivity occurs
in high-$T_c$ superconductors may be the extremely short coherence
length, $\xi$, about 15 {\AA} in the ab-plane.

For low-$T_c$ classic and layered magnetic superconductors, the
unusual behavior of the upper critical field near and below $T_N$
has been widely explored to verify pair-breaking interactions that
arise when magnetic and superconducting subsystems coexist
\cite{{Fischer'1978},{Matsumoto'1983},{Shrivastava'1984},{Schmidt'1996}}.
For high-$T_c$ superconductors, the upper critical fields at
temperatures where the magnetic order appears are generally unavailable
in the laboratory \cite{{Welp'1989},{Nakao'1988},{Rogacki'1988}}.
In this work, we discuss some low field effects that allow us to study
the interaction between antiferromagnetism and superconductivity in
high-$T_c$ materials.

\section{Unusual vortex dynamics in clastic antiferromagnetic
superconductors} \label{Unusual} The temperature dependence of the
lower critical field, $H_{c1}(T)$, was studied close to $T_N$ in
GdMo$_6$Se$_8$ \cite{Rogacki'1988a}.  The obtained results show
the vortex dynamics of isolated vortex lines during the flux first
penetration process.  The most striking feature of the $H_{c1}(T)$
curve is a drastic dip of $H_{c1}$ observed at $T_N$.  This
irregularity is a simple consequence of the pair-braking effect
caused by fluctuations enhanced near a phase transition. Similar
unusual behavior of $H_{c1}(T)$ is expected for high-$T_c$
antiferromagnetic superconductors in the case when the conduction
and magnetic electrons interact.
\begin{figure}[!htb]
\begin{center}
\includegraphics*[width=0.5\textwidth]{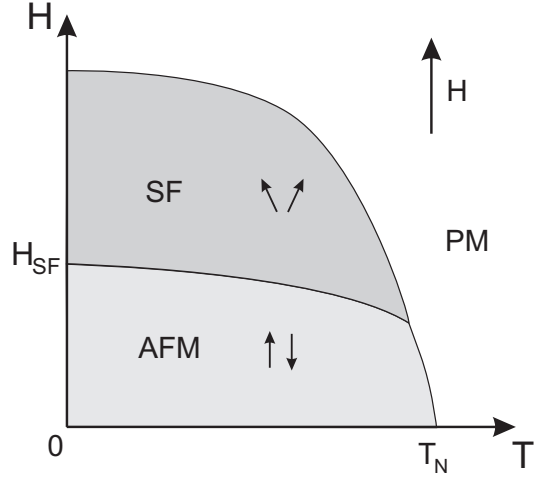}
\end{center}
\caption{Phase diagram of the two-sublattice antiferromagnet
with the easy axis oriented parallel to the external field
direction.  $H_{SF}$ is the field at which the transition from the
antiferromagnetic (AFM) to the spin-flop (SF) phase occurs.}
\label{H-T,AFM}
\end{figure}
\begin{figure}[!htb]
\begin{center}
\includegraphics*[width=0.48\textwidth]{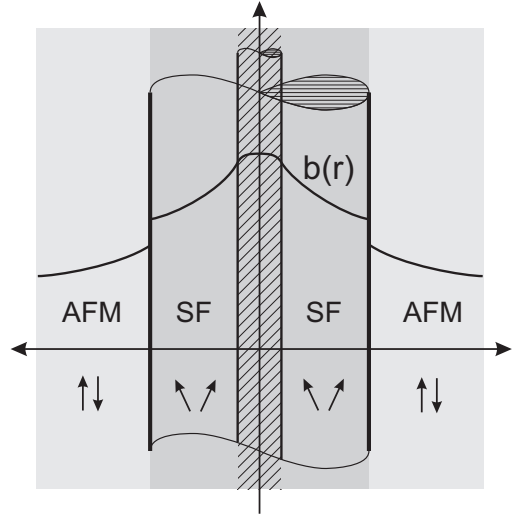}
\end{center}
\caption{Magnetic structure of the isolated vortex and the
distribution of magnetic induction ($b(r)$) around the vortex core
in the spin-flop (SF) and antiferromagnetic (AFM) phases
\cite{{Krzyszton'1980},{Krzyszton'sub}}. Dashed area illustrates
the vortex core.} \label{nic}
\end{figure}

Flux penetration was considered in detail for
classic magnetic superconductors when the magnetic
subsystem is a two-sublattice antiferromagnet with the easy axis
oriented parallel to the external field
\cite{{Krzyszton'1980},{Krzyszton'1984}}. Fig.~\ref{H-T,AFM} shows
the phase diagram of the two-sublattice aniferromagnet, where the
antiferromagnetic (AFM), spin-flop (SF), and paramagnetic (PM)
phases are present depending on the magnetic field and
temperature. This property requires the vortex line to have a
special magnetic structure when the field in the vortex core is
larger than the SF transition field, $H_{SF}$. When the density of
vortices increases, the spin-flop phase enlarges and the
individual vortex may have a magnetic structure as presented in
Fig.~\ref{nic}.  Consider the flux penetration process into
the AFM superconductor in the case when the vortices with the
described structure appear for $H$ close to $H_{c1}$.  It has been
predicted that such vortices have a new surface barrier for
penetration \cite{Krzyszton'1984}.  Thus, the flux penetration
process proceeds in two steps, as shown in Fig.~\ref{B(H),M(H)}.
\begin{figure}[!htb]
\begin{center}
\includegraphics*[width=0.4\textwidth]{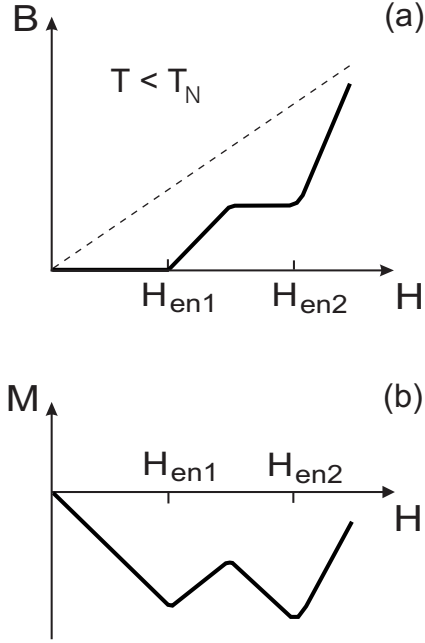}
\end{center}
\caption{Two-step flux penetration expected for the virgin
magnetization process of an antiferromagnetic superconductor.  The
$B(H)$ dependence (a) is transformed to the $M(H)$ dependence (b).
The magnetic easy axis of the superconductor is parallel to the
external field direction.} \label{B(H),M(H)}
\end{figure}
With increasing $H$ (see Fig.~\ref{B(H),M(H)}a), the Meissner
effect ($B = 0$) is observed until $H$ reaches the value of the
first penetration field, $H_{en1}$. Then, the usual flux penetration
process begins and $B$ increases. Next, when the field in the
vortex core is higher than $H_{SF}$, the SF phase appears and the
new shielding ($B =$ const) should be observed. For $H$ increased
above the characteristic value $H_{en2}$ (the second penetration
field), the magnetic flux penetrates the superconductor again.
The $B(H)$ dependence can be transformed to the $M(H)$ dependence
which can be easily measured at constant temperature.
Fig.~\ref{B(H),M(H)}b shows the $M(H)$ virgin curve for the two-step
flux penetration process expected for an AFM superconductor
magnetized in $H$ oriented parallel to the magnetic easy axis.

\begin{figure}[!htb]
\begin{center}
\includegraphics*[width=0.55\textwidth]{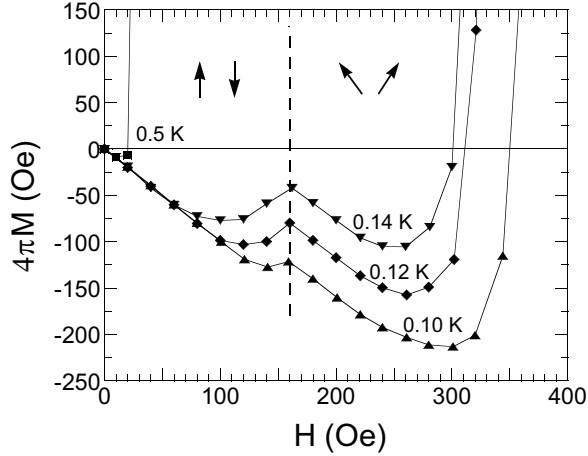}
\end{center}
\caption{Magnetization as a function of applied field for the
virgin state of the DyMo$_6$S$_8$ single crystal with the magnetic
easy axis parallel to the field direction.  As predicted, two-step
flux penetration is observed below $T_N = 0.4$~K. The arrows show
the possible configuration of the spin of Dy ions in the vortex
core.~\cite{Rogacki'2001}} \label{M(H)exp}
\end{figure}

The single crystal of DyMo$_6$S$_8$ was studied by magnetization
measurements below and above $T_N = 0.40$~K \cite{Rogacki'2001}.
The single crystal was a cube with the edge dimension of 0.2~mm
and a mass $m\simeq0.05$~mg.  It was oriented with the magnetic
easy axis (the [111] crystallographic triple axis) parallel to the
applied magnetic field, $H$.  For that orientation, the demagnetizing
factor was assumed to be $k=1/3$. Fig.~\ref{M(H)exp} shows the
unusual flux dynamics caused by the appearance of the SF phase in
the superconducting state. Here, magnetization behaves in the way
as predicted and illustrated in Fig.~\ref{B(H),M(H)}b. Two-step flux
penetration is clearly observed at $T < T_N$ and $H_{en2}(0.1$~K$)
\simeq 310$~Oe is found after correction for demagnetizing
effects.

The two-step flux penetration can be analyzed in the framework of
phenomenological theory with the free energy of an AFM
superconductor expressed in the following form \cite{Krzyszton'sub}:
\begin{equation}
F=\int dv \{ f_{S}+f_{M}+\frac{1}{8 \pi}( \mathbf{b}-4
\pi\mathbf{M}) ^{2} \},  \label{eq1}
\end{equation}
where $\mathbf{b}$ is the microscopic magnetic induction and
$\mathbf{M}= \mathbf{M}_{1}+\mathbf{M}_{2}$ is the magnetization
of a two-sublattice antiferromagnet.  The superconducting component
is introduced in the Ginzburg-Landau (GL) form:
\begin{equation}
f_{S}=\frac{\hbar ^{2}}{2m}\left| \left( \mathbf{\nabla
}-\frac{2ie} {c \hbar }\mathbf{a}\right)\Psi \right| ^{2}+\alpha
\left| \Psi \right| ^{2}+ \frac{1}{2} \beta \left |\Psi \right|
^{4}, \label{eq2}
\end{equation}
where $e$ and $m$ denote the charge and mass of an electron,
respectively, $c$ is the velocity of light, $\mathbf{a}$ is the
vector potential, $\Psi$ is the superconducting order parameter,
and $\alpha$ and $\beta$ are some phenomenological expansion
coefficients. The AFM energy is given by the following expression:
\begin{equation}
f_{M}=J\mathbf{M}_{1}\cdot\mathbf{M}_{2}+K\sum\limits_{i=1}^{2}\left(
M_{i}^{z}\right)^{2}
- \left| \gamma \right|\sum\limits_{i=1}^{2}\sum\limits_{j=x,y,z}(
\mathbf{\nabla } M_{i}^{j}) ^{2},  \label{eq3}
\end{equation}
where $J$ is the exchange constant between two AFM sublatices, $K$
denotes the single ion anisotropy constant, and $\sqrt{\left|
\gamma \right|}$ is the magnetic stiffness length.  According to
experiments, the so-called $s$-$f$ exchange interaction is weak
\cite{Matsumoto'1983}. This means that superconductivity ($\Psi$)
and magnetism ($\mathbf{M}$) interact via the electromagnetic
interaction and the order parameters are coupled via the vector
potential $\mathbf{a}$:
\begin{eqnarray}
&&\mathbf{b} = \nabla\times \mathbf{a}~, \nonumber \\
&&\mathbf{j}_{s}(\Psi^{+},\Psi) = \frac{c}{4\pi}\nabla\times
\left( \mathbf{b}-4 \pi \mathbf{M}\right).  \label{eq4}
\end{eqnarray}
The equilibrium conditions of the whole system can now be obtained
by minimization of the Gibbs free energy functional,
\begin{equation}
G=F-\displaystyle\frac{1}{4\pi}\int\mathbf{b}\cdot(\mathbf{b}-4
\pi \mathbf{M})dv~, \label{eq5}
\end{equation}
with respect to $\Psi$, $\mathbf{a}$, and $\mathbf{M}$.  Then, an
expression for $H_{en2}(B)$ is obtained as a final result
\cite{Krzyszton'sub}
\begin{equation}
H_{en2}(B)=\sqrt{B^{2}+H_{en2}^{2}(0)}~,  \label{eq6}
\end{equation}
where:
\begin{eqnarray}
&&2H_{en2}(0)=\frac{H_{SF}}{\displaystyle\sqrt{\frac{\varphi
_{0}}{\pi \lambda ^{2}B_{SF}}} \ln \left(\frac{\pi \lambda
^{2}B_{SF}}{\varphi _{0}}\right)}~, \nonumber \\
&&H_{SF}=2H_{c1}+z\frac{\varphi _{0}}{2\pi \lambda
^{2}}K_{0}\left(\frac{d}{\lambda}\right). \label{eq7}
\end{eqnarray}
Here, $\varphi_{0}$ is a flux quantum, $\lambda$ is a penetration
depth, $z$ is a coordination number, $K_0$ is the modified Bessel
function, and $d$ is a distance between the nearest vortices.
$H_{SF}$ is a thermodynamic critical field at which the transition
to the SF phase appears.

The experimental and calculated results obtained for DyMo$_6$S$_8$
are shown in Table~\ref{tab1}.  They agree well and this seems to
support the assumption that the dominant interaction between
superconducting and magnetic subsystems is the electromagnetic one,
even at low magnetic fields.  This assumption is sufficient to
describe quantitatively the anomalous virgin magnetization curves
observed in the AFM superconducting state.
\begin{table}[h]
\caption{The experimental (exp) and calculated (cal) second
penetration field, $H_{en2}$.  The experimental values were
obtained from Fig. \ref{M(H)exp} after correction for
demagnetizing effects.} \label{tab1}

\begin{tabular}{l|ccccc}
\hline
$T$~[K] &  $H_{en2}$(exp)~[Oe] & $H_{en2}$(cal)~[Oe]\\
 \hline
0.14    &  260                 & 215                \\
0.12    &  290                 & 240                \\
0.10    &  310                 & 265                \\
\hline
\end{tabular}
\end{table}

The two-step flux penetration process was used to estimate the
number of magnetic ions in the vortex core, and then, $\xi$ and
$\kappa$ (the GL parameter) in the AFM superconducting state
\cite{Rogacki'2001}.  The significant reduction of $\lambda$
was observed leading to a strong compression of the quantized flux
and resulting in a considerable decrease of $\kappa$ from 11,
obtained for $T$ just above $T_N$, to about 2.5 for $T$ below $T_N$.
The large reduction of $\kappa$ provides evidence that on decreasing
temperature the AFM superconductor tends to transform from a
type-II to a type-I superconductor, as predicted theoretically in
\cite{{Tachiki'1979},{Matsumoto'1983a}} and confirmed
experimentally for the ferromagnetic superconductor ErRh$_4$B$_4$
\cite{Gray'1983} .

\section{Interaction between antiferromagnetism and
superconductivity in selected high-$T_c$ materials}
\label{Antiferromagnetism}
Polycrystalline Gd123 was chosen to obtain $H_{c1}$ as a function
of $T$ close to $T_N = 2.2$~K to investigate the possible
interaction between the antiferromagnetic and superconducting
subsystems \cite{Rogacki'1992}.  Monotonic dependence of $H_{c1}(T)$
was observed and interpreted as evidence that no pair breaking is
present and that the AFM subsystem is effectively screened by
superconductivity.  This has been confirmed by ac susceptibility,
$\chi_{ac}$', measured as a function of $T$ with an ac field
$h_{ac}\leq10$~Oe at 200~Hz that has not revealed any AFM peak at
$T_N$.

The two-step flux penetration in layered high-$T_c$ superconductors
was considered in detail theoretically in \cite{Krzyszton'1994}.
Two cases were examined.  In the first, the external field was
applied parallel to both the CuO$_2$ layers and the magnetic easy
axis.  In the second, the external field was parallel to the layers
but perpendicular to the easy axis.  The calculations showed that
in the first case the vortex line has a complex magnetic structure
for $H_{en1}<\frac{1}{2}H_{SF}$ and the external field exceeds the
value $\frac{1}{2}H_{SF}$.  This structure consists of a SF domain
which is first created along the vortex core, like in classic
magnetic superconductors (see Fig.~\ref{nic}).  The appearance of
this structure may have a profound effect on both classic and
quantum flux creep in AFM high-$T_c$ superconductors
\cite{Krzyszton'2000}.

Dy123 single crystals with $T_c = 88$~K and $T_N = 0.9$~K were
used to verify the two-step flux penetration process in magnetic
high-$T_c$ superconductors \cite{Rogacki'1995}.
Magnetization virgin curves $M(H)$ were studied with a vibrating
reed technique at $T<T_N$ and $H$ parallel to the $c$-axis.  For
fully oxygenated Dy123, the $c$-axis is parallel to the magnetic
easy axis \cite{{Fischer'1988},{Clinton'1991}}, thus the
sample-field orientation fulfills the requirement to observe
two-step flux penetration \cite{Krzyszton'1994}.  No anomalous
behavior of $M$ was found for the field value for which the
transition from the AFM to the SF phase occurs. $M(H)$ curves
follow the analytical result for a type-II superconducting stripe
in a perpendicular magnetic field. This suggests a weak
interaction between superconducting and localized magnetic
electrons and provides verification that the existing reorientation
in the magnetic subsystem is externally shielded and can
not be observed by magnetic or transport measurements. In that
case, the possible interaction between antiferromagnetism and
superconductivity may be expected to be revealed when
antiferromagnetism is enhanced and/or superconductivity diminished.
There are several ways to lower $T_c$ in the Y123 type
superconductors.  One way is to reduce the oxygen content much
below 7 in a stoichiometric compound
\cite{{Jorgensen'1990},{Graf'1990},{Krekels'1992}}. Another is
substitution of RE for Ba
\cite{{Li'1988},{Wada'1989},{Woerden'1990},{Takita'1992}} or Sr
for Ba \cite{{Veal,'1989},{Dabrowski'1997},{Rogacki'2000}}.  Yet
another uses a weak Josephson coupling between grains in ultrathin
Dy123 films \cite{Beauchamp'1994}.  For these films, a clear peak
was observed at $T_N$ for resistance measured as a function of
temperature.  The peak was interpreted
as a result of the reduction of the intergranular Josephson
coupling by pair breaking due to enhanced intragranular
spin-disorder scattering at $T_N$.

\begin{figure}[!htb]
\begin{center}
\includegraphics*[width=0.57\textwidth]{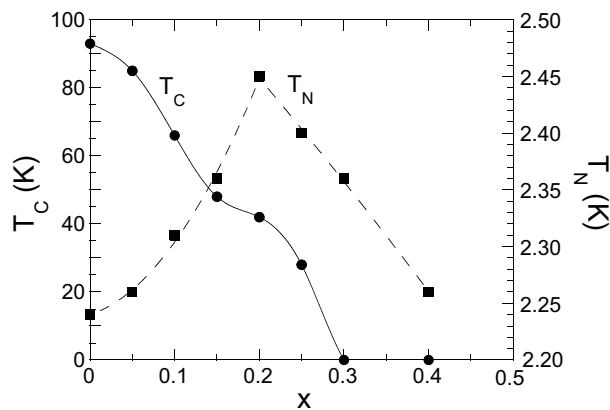}
\end{center}
\caption{Superconducting transition temperature, $T_c$, and
antiferromagnetic order temperature, $T_N$, versus composition $x$
for oxygen annealed Gd$_{1+x}$Ba$_{2-x}$Cu$_3$O$_{7-\delta}$.}
\label{Tc(x)}
\end{figure}

Samples of Gd$_{1+x}$Ba$_{2-x}$Cu$_3$O$_{7-\delta}$ solid solution
were synthesized resulting in superconductors with $T_c$
decreasing from 93~K to about 40~K and $T_N$ increasing from 2.24
to 2.45~K for $x$ changing from 0 to 0.2. The final heat treatment
was performed for 24 hours at 400 $^o$C in oxygen and 700 $^o$C in
Ar to obtain superconducting and non-superconducting samples,
respectively. For the superconducting samples, the maximized
oxygen content increased from 6.93 to 7.03 for substitution levels
varying from $x = 0$ to $x = 0.2$.  Details about the preparation
procedure were published elsewhere \cite{Plackowski'1995}.
Fig.~\ref{Tc(x)} shows $T_c(x)$ and $T_N(x)$ dependencies that
reveal the composition with $x = 0.2$ as optimal for our purpose.
For samples with $x > 0.2$, $T_N$ decreases and the peak observed
in specific heat measurements at $T_N$ gets smaller and wider
indicating that the Gd ions may not be perfectly magnetically
ordered and, quite probably, they are also not homogenously
distributed.

\begin{figure}[!htb]
\begin{center}
\includegraphics*[width=0.55\textwidth]{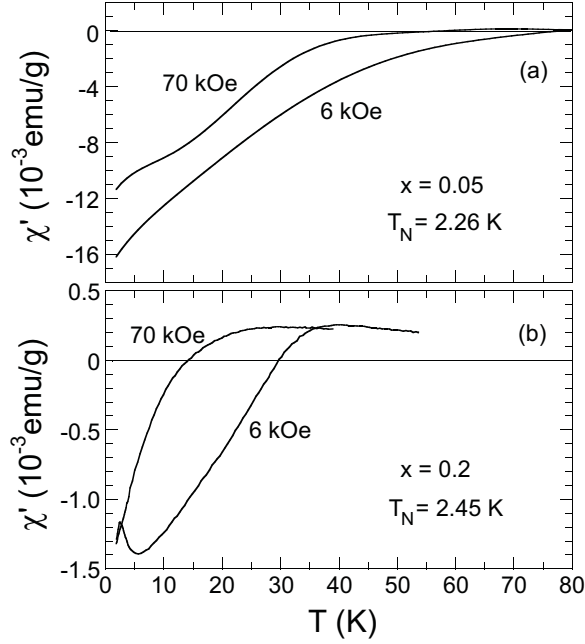}
\end{center}
\caption{Real part of the ac susceptibility, $\chi$', for the
oxygen annealed Gd$_{1+x}$Ba$_{2-x}$Cu$_3$O$_{7-\delta}$ samples
with $x = 0.05$ (a) and 0.2 (b) measured at $H = 6$ and 70~kOe.
Clear anomaly is present at $T_N$ for the $x = 0.2$ sample
measured at $H = 6$~kOe.} \label{X(T)}
\end{figure}

Fig.~\ref{X(T)} shows the $\chi_{ac}$'$(T)$ results obtained for
the oxygen annealed samples with $x = 0.05$ (Fig.~\ref{X(T)}a) and
0.2 (Fig.~\ref{X(T)}b). A pronounced peak at $T_N$ is observed at
$H = 6$~kOe for the $x = 0.2$ sample. This peak is interpreted
as evidence that the magnetic and superconducting subsystems
interact and superconductivity is not able to screen the AFM
fluctuations being enhanced close to the phase transition.  We
believe that the pair breaking effect is a simple reason for the
observed peak.  This effect can not be studied in the usual way
because $H_{c2}$ is still to high to be measured at temperatures
close to $T_N$ for the sample with $x = 0.2$.
\begin{figure}[!htb]
\begin{center}
\includegraphics*[width=0.55\textwidth]{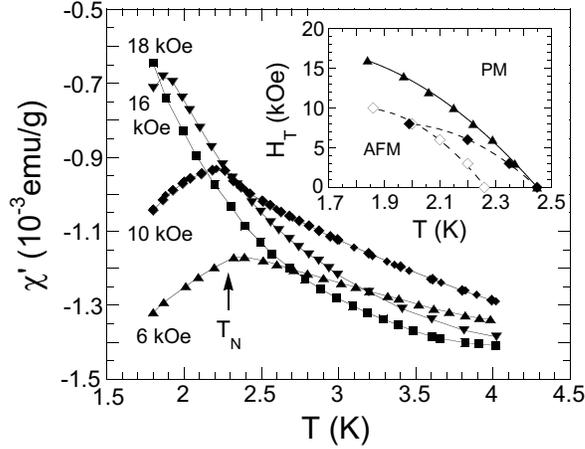}
\end{center}
\caption{Low temperature part of the ac susceptibility, $\chi$',
for the oxygen annealed Gd$_{1+x}$Ba$_{2-x}$Cu$_3$O$_{7-\delta}$
sample with $x = 0.2$ measured at several applied fields. The
inset shows the $H_T$-$T$ phase diagram for the superconducting
(triangles, solid line) and non-superconducting (diamonds, broken
line) samples with $x = 0.05$ (open symbols) and 0.2 (solid
symbols).} \label{X(T)atTn}
\end{figure}

Fig.~\ref{X(T)atTn} shows the low temperature part of
$\chi_{ac}$'$(T)$ obtained for the oxygen annealed $x = 0.2$ sample
at several magnetic fields.  Along with increasing $H$, the AFM
peak shifts to lower temperatures, as expected.  This observation
is consistent with the results obtained for the specific heat
measurements of the Gd123 non-substituted compound
\cite{Meulen'1988}. Results presented in Fig.~\ref{X(T)atTn} are
used to construct the $H_T$-$T$ phase diagram, where $H_T$ is a
field for which the transition from the AFM (or SF) to the PM
phase occurs.  Similar $\chi_{ac}$'$(T)$ measurements have been
performed for the non-superconducting $x = 0.2$ sample annealed in
Ar atmosphere.  The $H_T$-$T$ phase diagram is shown
in the inset of Fig.~\ref{X(T)atTn} for both the superconducting
and the non-superconducting sample.  A distinct difference between
the $H_T(T)$ dependencies is observed and this is important
evidence that the superconducting $x = 0.2$ sample is free of a
significant amount of the oxygen deficient non-superconducting
phase.  In Fig.~\ref{X(T)atTn}, the only single maximum is present
for every $\chi_{ac}$'$(T)$ curve, so we believe that the
$x = 0.2$ sample is homogenous and no separation of the magnetic
superconducting and the magnetic normal phases appears. Thus, the
$\chi_{ac}$' peak observed for the $x = 0.2$ sample at $T_N$ seems
to be a result of the interaction between coexisting
antiferromagnetic and superconducting subsystems.

\section{Conclusion}
Interaction between antiferromagnetism and superconductivity in
low-$T_c$ materials is usually studied by measuring $H_{c2}(T)$,
$H_{c1}(T)$, $\chi_{ac}$'$(T)$, and $M(H)$ which reveal anomalies at
$T_N$ and $H_{SF}$.  When the anomalous dependencies are known, the
type of interaction between the long-range antiferromagnetic order
and superconductivity may be determined.  In this work, we examined
the two-step flux penetration process for the initial
magnetization of the classic magnetic superconductor DyMo$_6$S$_8$.
This unusual flux penetration was interpreted as the consequence of
the spin-flop phase appearing in the vortex core.  Based on that
observation and analyzing the free energy of a magnetic
superconductor, the second penetration field for the two-step flux
penetration, the superconducting coherence length, and the
Ginzburg-Landau parameter were estimated. It was also shown
that for the low-$T_c$ superconductor DyMo$_6$S$_8$ the interaction
between magnetic and conduction electrons is mostly electromagnetic.
In the case of high-$T_c$ superconductors, some anomalies expected
in the $H_{c1}(T)$, $M(H)$, and $\chi_{ac}$'$(T)$ dependencies were
described for GdBa$_2$Cu$_3$O$_{7-\delta}$,
DyBa$_2$Cu$_3$O$_{7-\delta}$, and
Gd$_{1+x}$Ba$_{2-x}$Cu$_3$O$_{7-\delta}$, respectively.  No unusual
properties were found for those compounds except where Gd is
partially substituted for Ba.  For this compound, with $x = 0.2$,
the interaction between antiferromagnetism and superconductivity
leads to a peak in the $T$ dependence of $\chi_{ac}$' at $T_N$.
This observation was interpreted as a result of the pair-breaking
effect that is present in a material where $T_N$ is increased and
$T_c$ is decreased compared to the non-substituted compound.  In
non-substituted or weakly substituted Gd123 the interaction between
magnetic and conduction electrons is very weak and the existing
reorientation of the magnetic moments of the Gd ions at $T_N$ is
fully screened by superconductivity.

\section*{Acknowledgements}
The part concerning low-$T_c$ superconductors mainly emerged from
work done with Drs. C. Su\l kowski, E. Tjukanoff, and T.
Krzyszton. The part about high-$T_c$ materials mainly based on
work performed in close cooperation with Dr. Z. Bukowski, Prof. P.
Esquinazi, and Prof. B. Dabrowski. The author also would like to
acknowledge many stimulating discussions with T. Krzyszton and P.
Tekiel. This work was supported by the State Committee for
Scientific Research (KBN) within the Project \mbox{No. 2 P03B 125
19}.

\end{document}